\documentclass[aps,prl,showpacs,twocolumn,superscriptaddress]{revtex4}
\usepackage{graphicx}
\usepackage{amssymb}
\usepackage{dcolumn}
\usepackage{bm} 
\begin{document}
\title{Dominant mobility modulation by the electric field effect at the LaAlO$_3$ / SrTiO$_3$ interface}
\author{C. Bell}
\affiliation{Department of Advanced Materials Science, University of Tokyo, Kashiwa, Chiba 277-8651, Japan}
\affiliation{Japan Science and Technology Agency, Kawaguchi, 332-0012, Japan}
\author{S. Harashima}
\affiliation{Department of Advanced Materials Science, University of Tokyo, Kashiwa, Chiba 277-8651, Japan}
\author{Y. Kozuka}
\affiliation{Department of Advanced Materials Science, University of Tokyo, Kashiwa, Chiba 277-8651, Japan}
\author{M. Kim}
\affiliation{Department of Advanced Materials Science, University of Tokyo, Kashiwa, Chiba 277-8651, Japan}
\author{B. G. Kim}
\affiliation{Department of Advanced Materials Science, University of Tokyo, Kashiwa, Chiba 277-8651, Japan}
\affiliation{Department of Physics, Pusan National University, Busan 609-735, Korea}
\author{Y. Hikita}
\affiliation{Department of Advanced Materials Science, University of Tokyo, Kashiwa, Chiba 277-8651, Japan}
\author{H. Y. Hwang}
\affiliation{Department of Advanced Materials Science, University of Tokyo, Kashiwa, Chiba 277-8651, Japan}
\affiliation{Japan Science and Technology Agency, Kawaguchi, 332-0012, Japan}
\date{\today}

\newcommand{\sto}{SrTiO$_3$}          
\newcommand{\stos}{SrTiO$_3$ }          
\newcommand{\stod}{SrTiO$_{3-\delta}$}
\newcommand{\stods}{SrTiO$_{3-\delta}$ }
\newcommand{\lao}{LaAlO$_3$}         
\newcommand{\laos}{LaAlO$_3$ }         
\newcommand{\etal}{{\it et al$.$ }}     
\newcommand{\etaln}{{\it et al$.$}}   
           
\begin{abstract}Caviglia {\it et al$.$} [Nature (London) {\bf 456}, 624 (2008)] have found that the superconducting LaAlO$_3$/SrTiO$_3$ interface can be gate modulated. A central issue is to determine the principal effect of the applied electric field. Using magnetotransport studies of a gated structure, we find that the mobility variation is almost five times as large as the sheet carrier density. Furthermore, superconductivity can be suppressed at both positive and negative gate bias. These results indicate that the relative disorder strength strongly increases across the superconductor-insulator transition.
\end{abstract}

\maketitle

The strength of the electric field effect (EFE) in accumulating or depleting carriers in a conducting channel is central not only to many semiconductor devices found ubiquituously in modern electronics, but also in current research into achieving novel physics using tunable materials. Complex oxides are one case where the electronic ground state of the system is highly sensitive to the carrier density \cite{ahn_rmp2006,caviglia_nature2008}. Among the commonly studied oxide materials, \stos has attracted much attention due to its high electron mobility and electric permittivity at low temperatures, which facilitates large electric field effects \cite{parendo_prl2005}. Many recent oxide EFE devices have utilized \stos substrates as a crucial component of the experiment: both the metallicity and superconductivity of \stos have been modulated \cite{nakamura_apl2006,thiel_science2006,ueno_natmat2008,caviglia_nature2008}.

Recently Caviglia \etal \cite{caviglia_nature2008} strikingly demonstrated that the EFE could be used to modulate the superconductivity which appears \cite{reyren_science2007} in the metallic gas formed between the two insulators \laos and \sto. Since its first discovery \cite{ohtomo_nature2004} the origin and physics of this metallic layer has been intensively investigated. Room temperature scanning electron energy-loss spectroscopy and conducting scanning probe measurements have set an upper limit of $\sim$7 nm for the gas thickness in annealed or high pressure grown samples \cite{nakagawa_natmat2006,basletic_natmat2008}. However it is still unclear how the gas thickness is correlated to the sheet resistance at low temperatures, which itself can be changed by several orders of magnitude with oxygen pressure during \laos growth \cite{ ohtomo_nature2004,brinkman_natmat2007}, or the EFE \cite{thiel_science2006, caviglia_nature2008}. 

In this context, Caviglia \etal assumed that the superconductivity suppression by the EFE is due to the reduction of the carrier density of the electron gas. However it was unclear how the mobility of the electron gas changed in these initial EFE experiments, since it was not probed. A direct measurement of the mobility is vital in order to experimentally determine what is the effective tuning parameter of the superconductor-insulator transition. In this Letter, we describe a detailed study of the magnetotransport properties of a \lao/\stos interface, for which superconductivity can be fully suppressed by both positive and negative gate bias. Normal state magnetotransport measurements were also made above the upper critical field at which superconductivity is destroyed. From these data we find that not only does the carrier density vary with applied gate voltage, but a significant change of the electron mobility also occurs, which in fact the dominates the change in the normal state conductivity. These results are crucial for a full understanding of the nature of the transition from the superconducting to insulating state in this system, and others like it which utilize \stos as an active element \cite{parendo_prl2005,nakamura_apl2006,thiel_science2006,parendo_prb2006,matthey_prl2007,salluzzo_prl2008,ueno_natmat2008}. 

Our sample was grown by pulsed laser deposition as described elsewhere \cite{bell_apl2009} using an oxygen pressure of $1.33 \times 10^{-3}$ Pa. The \laos thickness was 10 unit cells, as monitored using {\it in-situ} reflection high-energy electron diffraction. The substrate was 5 mm $\times$ 5 mm $\times$ 0.5 mm \stos (100) with a TiO$_2$ terminated surface. Using optical lithography, a six contact Hall bar (central bar length 500 $\mu$m, width 100 $\mu$m) was patterned onto the \stos utilizing amorphous AlO$_{\mathrm{x}}$ as a hard mask, which was lift-off patterned at room temperature prior to the \laos growth.

The Hall bar was ultrasonically wirebonded with Al wire to form Ohmic contacts, and the gate contact was made on the back of the \stos substrate via conducting silver epoxy. A quasi d$.$c$.$ bias current sweep between $1-100$ nA was used for all transport measurements. The back gate leakage current was $< 0.1$ nA for all temperature and voltages. Sheet resistance versus temperature, $R(T)$, was measured in the temperature range 0.05 K $\le T\le0.5$ K in a dilution refrigerator with a base temperature of 10 mK. The gate voltage $V_g$ was first swept between the maximum and minimum values of $\pm$100 V to remove history effects \cite{caviglia_nature2008}, before measurements were made at 25 V intervals. Similar measurements were also performed in a helium-4 cryostat for $2$ K $\le T \le 300$ K, which was more convenient to investigate thermal history effects associated with changes in $V_g$. \begin{figure}[h]\includegraphics[width=8.5cm]{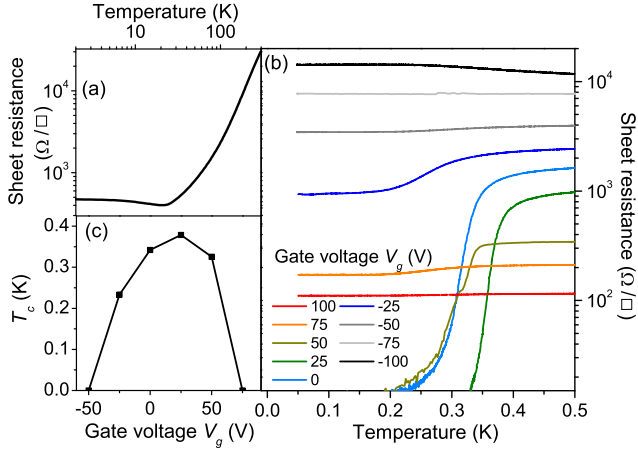}\caption{\label{fig1}(Color) (a) $R(T)$ from $T=300$ K before biasing. (b) Low temperature $R(T)$ for various gate voltages. (c) Superconducting critical temperature $T_c$ versus gate voltage.}\end{figure}

Figure \ref{fig1}(a) shows $R(T)$ in the high temperature regime before applying a gate voltage, showing a slight upturn around $T=20$ K, similar to that observed elsewhere \cite{ohtomo_nature2004,brinkman_natmat2007,reyren_science2007}. Figure \ref{fig1}(b) shows the low temperature $R(T)$ for various $V_g$ applied. A clear systematic increase of the resistance is observed for $V_g <0$ and decrease for $V_g >0$. Superconductivity is found for the range -50 V $< V_g < 75$ V, with $T_c = 378$ mK at $V_g = +25$ V, as shown in the phase diagram of Fig$.$ 1(c). Here for simplicity we have defined $T_c$ as the temperature at which the sheet resistance falls below 50\% of the value at $T = 0.5$ K. Similar to Ref$.$ \cite{caviglia_nature2008}, we can completely suppress $T_c$ by applying a negative gate voltage (removing electrons), but in this sample we can also add electrons (positive $V_g$) and reduce $T_c$ to zero. 
\begin{figure}[h]\includegraphics[width=8.5cm]{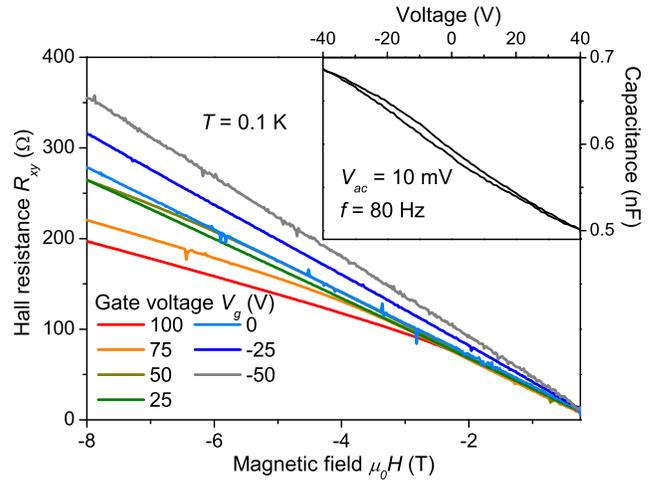}\caption{\label{fig2}(Color) Antisymmetrized Hall resistance versus magnetic field, $R_{xy}(H)$ for various $V_g$ at $T=0.1$ K. Lines are guides to the eye. Inset: Capacitance versus voltage at $T = 0.1$ K.}\end{figure}

\begin{figure}[h]\includegraphics[width=8.5cm]{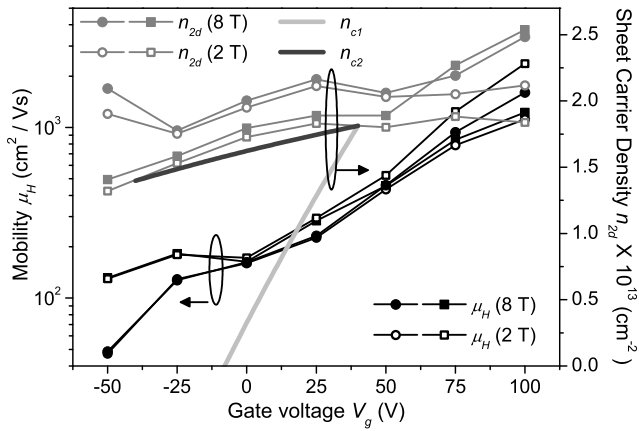}\caption{\label{fig3}Sheet carrier density $n_{2d}$ and electron mobility $\mu_H$ versus $V_g$, at $T = 0.1$ K (squares) and $T=2$ K (circles). Closed and open symbols refer to high field (8 T) and low (2 T) field values of the Hall coefficient respectively. Heavy lines are carrier density changes estimated from the capacitance data using top ($n_{c1}$) and bottom ($n_{c2}$) areas. Other lines are guides to the eye.}\end{figure}

The normal state magnetotransport properties were measured above the superconducting upper critical field, $H_{c2}$, using a magnetic field $\mu_0H \gg  150$ mT $> \mu_0 H_{c2}$. The Hall resistance data at $T = 0.1$ K are shown in Fig$.$ \ref{fig2}. The extracted sheet carrier density, $n_{2d}\sim1.8 \times 10^{13}$ cm$^{-2}$ for $V_g=0$ V at $\mu_0H = 2$ T is comparable to other studies. Applying positive $V_g$, $R_{xy}(H)$ develops a non-linearity, hence we extract the sheet carrier density from the Hall coefficient at both 8 T and 2 T. At $V_g = -75$ V and $-100$ V $R_{xy}(H)$ could not be measured reliably, suggesting that inhomogeneities develop in the wire at higher sheet resistances. The carrier density modulation for both high (8 T) and low (2 T) field fits show the same trend with $V_g$: a reduction of $\sim45$ $\%$ between $V_g=+100$ V and $-50$ V (Fig$.$ \ref{fig3}). The modulation of the sheet carrier density in this voltage range is linear, meaning that the device is operating as a conventional metal oxide field effect transistor. Measurements at $T = 2$ K show a similar trend, with only a slight difference in the value of $n_{2d}$. 

The relative change in charge with $V_g$ can also be estimated using the integrated capacitance versus voltage ($C(V_g)$, inset Fig$.$ \ref{fig2}). However we found that the charge variation, as scaled by the area of the Hall bar was significantly larger than that measured by the Hall effect, ($n_{c1}$ data in Fig$.$ \ref{fig3}). We find better agreement with the Hall effect by scaling with the bottom gate area ($n_{c2}$, Fig$.$ \ref{fig3}), however this would imply some conduction beneath the AlO$_x$ hard mask, despite the large measured resistance $> 100$ G$\Omega / \Box$. The clear hysteresis observed in the $C(V_g)$ data (inset Fig$.$ \ref{fig2}), suggests the presence of weak induced interfacial ferroelectricity due to the presence of a large electric field \cite{saifi_prb1970, neville_jap1972}. Thus charge trapping is a more likely scenario for the overestimate of the $n_{2d}$ change from the $C(V_g)$ data, and we rely on the Hall data as a direct probe of the free carrier density in the electron gas. 

Next we focus on the change in electron mobility, which is independent of the weak Hall effect non-linearity. Using the standard Hall mobility formula $\mu_H = (en_{2d}R)^{-1}$ where $e$ is the electronic charge, $\mu_H(V_g)$ was calculated. The $\mu_H$ data are shown in Fig$.$ \ref{fig3}, together with $n_{2d}(V_g)$. These results show a clear and consistent reduction of the electron mobility as $V_g$ decreases from $100$ V to $-50$ V, at $T = 100$ mK. This change is a factor of 9.3 (8 T Hall coefficient data), with a similar change at $T = 2$ K. Thus in this system the change in conductivity is dominated by the mobility change, and not the sheet carrier density, which varies by a factor of only $\sim 1.8$ in the same $V_g$ range. At even higher temperature ($T = 20$ K, data not shown) we find the same dominance of the mobility change, emphasizing the robustness of this result. 

We have modeled the carrier distribution assuming a triangular well with the \laos acting as an infinite potential barrier at the origin \cite{stern_prb1972}. We use the band structure assumptions of Ueno \etal \cite{ueno_natmat2008}, i$.$e$.$ that the three doubly-degenerated conduction band valleys centered at the $\Gamma$ point of \stos show no band-splitting, and consists of one heavier mass band (effective mass of $m_h^{\star} = 4.8 m_0$, where $m_0$ is the bare mass) and two lighter bands ($m_l^{\star} =  1.2 m_0$). At $V_g = 0$, we take a self consistent average electric field as the confining potential. This is non-linear in the carrier density, as given by $E_{av} = A ( \exp (0.5eBn_{2d} \varepsilon_0^{-1}) -1 )$, where $\varepsilon_0$ is the vacuum permittivity, $A = 8.349 \times 10^4$ V/m, and $B = 4.907\times10^{-10}$ m/V \cite{neville_jap1972}. Using the 2 K high field value of $n_{2d}= 2.0\times 10^{13}$ cm$^{-2}$, we find an effective relative permittivity $\varepsilon_r \sim 1.5\times 10^4$ and $E_{av} \sim 1.2 \times 10^5$ V/m. 

\begin{figure}[h]\includegraphics[width=8.5cm]{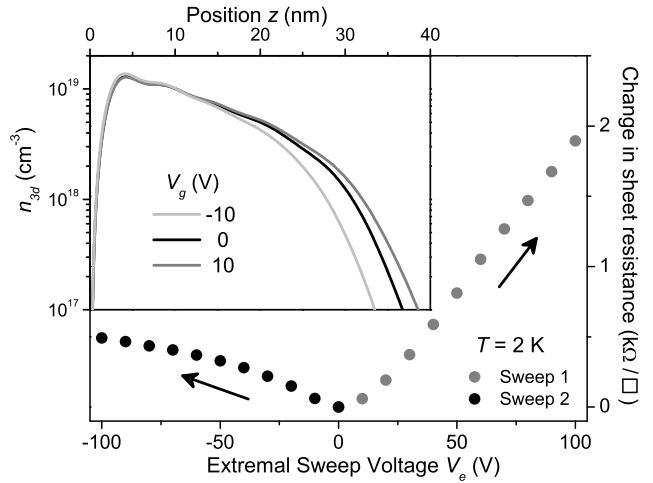}\caption{\label{fig4}Sheet resistance at $V_g = 0$ V and $T=2$ K after applying an extremal voltage $V_e$. Sweep two was performed after resetting the system at $T=300$ K. Inset: Electron distribution calculations at three bias voltages.}\end{figure}

In this approximation the solutions of the Schr{\"o}dinger equation are Airy functions of the form $ \zeta_i(z) = \mathrm{Ai}(z\alpha - [1.5\pi(i-0.25)]^{2/3})$, where $\alpha = (\hbar^2/2m^{\star}eE_{av})^{1/3}$, $z$ is the direction normal to the interface, and $i$ is an integer. Electrons are added progressively into the energy bands until the total charge $n_{2d}$ is reached, and the Fermi energy $E_F$ is determined self consistently. The electron distribution can then be calculated using the eigen energies $E_i$ and corresponding wavefunctions $\zeta_i$ using \begin{equation} n_{3d}(z) = \sum_{j = l,h} \left( \frac{g_j m_j^{\star}}{2\pi \hbar^2 } \sum_i (E_F - E_i)|\zeta_i(z)|^2 \right) \end{equation} with $g_h = 2, g_l = 4$ and $E_i = eE_{av}\alpha^{-1}[1.5\pi(i-0.25)]^{2/3}$. The result of this calculation gives an electron distribution as shown in the inset of Fig$.$ \ref{fig4}. The peak volume carrier density $n_{3d}^{\mathrm{max}} = 1.3 \times 10^{19}$ cm$^{-3}$, is well within the range of densities for which superconductivity in bulk \stos is found \cite{koonce_pr1967}.

To give a real space picture of the physics occuring when applying finite $V_g$, we must consider two effects. Firstly the insulating \stos substrate acts as simply a capacitor which moves charge to the interface. The electric field $E_{av}$ confining this charge will then change with $n_{2d}$ according to the previous non-linear equation. However at the same time the conduction and valence bands of the bulk \stos must be connected continuously with those of the metallic gas. Thus for $V_g <0$ ($V_g >0$) an additional compression (expansion) of the electron gas will occur due to band bending, similar to previous discussions in the case of mobility suppression in gated $n$-AlGaAs-GaAs heterojunctions \cite{hirakawa_prl1985}.  Qualitatively we can illustrate the effect of this bias using the above model in a small voltage range ($\pm 10$ V). We add the non-linear $E_{av}$ and the applied electric field due to $V_g$ to define the new confining potential \cite{stern_prb1972} and recalculate the electron distribution for $n_{2d}$ interpolated at $V_g = \pm10$ V. These data are also shown in the inset of Fig$.$ \ref{fig4}, and clearly demonstrate the compression (expansion) effect with negative (positive) gate bias. 

In this range, $V_g$ is a small perturbation on $E_{av}$. At larger $V_g$ a fully self-consistent calculation is necessary to incorporate the local non-linear permittivity $\varepsilon_r(V_g,z)$ \cite{saifi_prb1970, neville_jap1972}. Although very computationally demanding, the response of the electron distribution and lattice relaxation effects must also be included \cite{hamann_prb2006} in order to achieve a full quantitative understanding of the gating effect. The inclusion of non-linearities of the permittivity tend to collapse the value $\varepsilon_r$ close to the interface, leading to further compression the gas closer to the \laos / \stos interface \cite{susaki_prb2007}, and may even counter-intuitively increase $n_{3d}^{\mathrm{max}}$ at large negative biases. As noted above, the superconductivity in our sample shows a much more sensitive response to an applied voltage that the previous study, where $T_c$ could only be suppressed to zero only in the regime $V_g <0$. Also our maximum in $T_c$ is larger than that of Ref$.$ \cite{caviglia_nature2008}, despite our lower starting value of $n_{2d}$. This contrast is therefore not due simply to the different sheet carrier densities in the system, but rather it depends on the detailed density distribution $n_{3d}(z)$. This is critical for understanding the superconducting phase diagram, and is not universal in the presence of non-linearities in $\varepsilon_r$. 

The decrease of the mobility is thus correlated with the loss of the lower density `tail' region of the $n_{3d}(z)$ distribution, and an increased relative contribution of interface scattering at the \lao/\stos interface as the center of mass of the electron gas moves closer to the \lao. Additionally, any decrease in $\varepsilon_r$ will enhance the scattering cross-section of previously screened ionized impurities, and compound the mobility reduction of the electron gas. This result has important implications when discussing the suppression of the superconducting state since changes in the disorder must be considered in additional to the change in $n_{2d}$. In this sample, a relatively weak but clear non-linearity in the Hall effect measurement was found, (Fig$.$ \ref{fig2}). In brief, this Hall effect non-linearity can be caused by multiple parallel conduction paths with different electron mobilities, or possibly by magnetic contributions. The former cause is more likely in the general case, and would naturally arise due to the concomitant distribution of the mobility and electron density throughout the thickness of \sto. The reduction of the Hall effect non-linearity for larger negative gate voltages is then a natural consequence of the loss of the low carrier density, higher mobility tail of the electron distribution. The increased electric field squeezes the electron distribution further towards the \lao/\stos interface, homogenizing the mobility distribution. 

Finally we briefly discuss history effects associated with irreversible changes in the resistance at low temperatures as $V_g$ is swept. These changes can only be removed by warming to room temperature. The main panel Fig$.$ \ref{fig4} shows how the change in $R$ at $(V_g=0)$ behaves as a function of the previous extremal voltage applied. That is to say, we applied $V_g$ to a maximum positive (or negative) value of $V_e$, again set $V_g=0$ and then measured $R$. First $V_e$ was ramped from 0 to -100 V in 10 V steps. After this, the sample was warmed to 300 K to reset the system, and cooled again to 2 K, and $V_e$ was then ramped from 0 to +100 V. A clear asymmetry in $R(V_e)$ is found: positive $V_e$ induces a much larger differential increase in $R$ than negative $V_e$. Such an asymmetry may be understood via the discussion above concerning the distortion of the electron distribution by the gate voltage. Positive $V_g$, extending the electron distribution deeper into the \stos substrate, causes electrons to fall into previously unfilled trap states. After the gate voltage is removed, and when the emission rate from the trap states is low, the system does not return to the previous state, but at a higher resistance, as observed. For $V_g<0$ the electrons are pressed closer to the \laos layer, but a significant number of new traps are not revealed, consistent with the asymmetry of the resistance change shown in Fig$.$ \ref{fig4}. The increase in resistance for $V_g<0$ is then assigned to additional charge trapping due to interfacial ferroelectricity \cite{zuleeg_sse1966}, the presence of which is suggested by the capacitance data already discussed. 

In conclusion we have studied the EFE at the \lao/\stos interface, and have shown that the electron mobility plays a dominant role in controlling the conductivity of this system. These data are consistent with a distortion of the electron wavefunction towards the interface. Thus variations in the effective disorder may dominate the modulation of the superconducting transition, which in our case could be suppressed to $T_c = 0$ K using both positive and negative gate voltages. Moreover we expect that these changes in electron scattering will generally be present when using the EFE to tune any superconductor-insulator transition, evidence for which has been noted elsewhere \cite{parendo_prb2006,goldmanprivate}.

We thank M. Lippmaa for use of cleanroom facilities, and A. M. Goldman for useful discussions. CB acknowledges partial funding from the Canon Foundation in Europe.

\end{document}